\documentstyle[preprint,aps]{revtex}
\begin{document}
\draft
\preprint{
\parbox{4cm}{
\baselineskip=12pt
YCTP-P26-97\\
December, 1997\\
\hspace*{1cm}
}}
\title{Mass Generation Mechanism in Supersymmetric\\
       Composite Model with Three Generations}
\author{Noriaki Kitazawa\thanks{e-mail: kitazawa@zen.physics.yale.edu}}
\address{Department of Phyiscs, Yale University,
         New Haven, CT 06520, USA
         \thanks{On leave: Department of Physics,
                           Tokyo Metropolitan University,
                           Tokyo 192-03, Japan}}
\maketitle
\begin{abstract}
We propose a supersymmetric composite model with three generations
 in which supersymmetry and electroweak symmetry are broken dynamically,
 and masses of quarks and leptons are generated
 without introducing any mass scales by hand.
All the mass scales in the model
 are expected to be generated dynamically.
The mechanism to have mass hierarchy is explicitly described,
 although the roughly estimated mass spectrum of quarks and leptons
 does not exactly coincide with the realistic one.
\end{abstract}
\newpage

\section{Introduction}
\label{sec:intro}

Recently, many supersymmetric composite models
 have been proposed towards a better understanding
 of the generation structure, the mass hierarchy of quarks and leptons,
 and the supersymmetry and electroweak symmetry
 breaking\cite{N-S,L-M,L,K-O,H,K-L-S,H-O,O}.
In spite of these many efforts,
 we still do not have satisfying understanding.
In most of models,
 some mass scales must be introduced by hand
 to have higher dimensional (non-renormalizable) effective 
 interactions in the superpotential for quark mass generation.

In this paper we propose a supersymmetric composite model
 with three generations of quarks and leptons,
 in which there is no mass scale introduced by hand.
All the mass scales are expected to be generated dynamically.
Although the resultant mass hierarchy of quarks and leptons
 does not exactly coincide with the realistic one,
 the mechanism for the generation of the mass hierarchy itself
 might be true.

We begin with a brief review
 of the compositeness structure of one generation
 which comes from the model proposed by Nelson and Strassler\cite{N-S}.
Consider the following particle content:
\begin{center}
\begin{tabular}{ccc}
                   & \ $SU(2)_H$ \ & SU(5) \ \\
$P^a$              &   $2$         & $5$   \\
$N$                &   $2$         & $1$   \\
${\bar \Phi}_{1a}$ &   $1$         & $5^*$ \\
${\bar \Phi}_{2a}$ &   $1$         & \ $5^*$, \\
\end{tabular}
\end{center}
 where $SU(5) \supset SU(3)_C \times SU(2)_L \times U(1)_Y$
 contains the standard model gauge group
 and $SU(2)_H$ is the additional hypercolor gauge interaction
 which becomes strong at some high scale $\Lambda_H$
 and confinement is expected.
All fields are chiral superfields
 and $a$ is the index of the $SU(5)$ representation.
The representation is vector-like,
 if we consider only one simple gauge group,
 and the anomaly cancellation is trivial.
Although we use $SU(5)$ representations
 to describe the quantum number of the standard model gauge group
 throughout this paper,
 we do not assume the grand unification.
Using the technique developed by Seiberg et al.\cite{S},
 we can identify massless low energy effective fields as
\begin{equation}
 \Sigma^{ab}
  \sim [P^a P^b] \equiv \epsilon^{\alpha\beta} P^a_\alpha P^b_\beta,
 \quad
 \Phi^a \sim [P^a N],
\label{contraction}
\end{equation}
 where $\Sigma^{ab}$ and $\Phi^a$ are identified
 as the matter multiplet in \underline{10} representation
 and the Higgs multiplet in \underline{5} representation
 in the standard $SU(5)$ grand unified theory, respectively.
If we introduce the tree level superpotential
\begin{equation}
 W_{tree} = {\bar \Phi}_{2a} \eta^a_b [P^b N]
\end{equation}
 which is the general form in the given symmetry,
 the masses of Higgs particles are generated as
\begin{equation}
 W_{tree} \longrightarrow \Lambda_H {\bar \Phi}_{2a} \eta^a_b \Phi^b,
\end{equation}
 where $\Lambda_H$ is the scale of the hypercolor dynamics,
 and the explicit $SU(5)$ breaking effect is incorporated
 in the coupling matrix $\eta^a_b$.

It is expected that
 the non-perturbative hypercolor dynamics generates
 the superpotential\cite{S}.
If we consider the third generation, we can write it down as
\begin{equation}
 W_{dyn} = {1 \over {2^3}}
           \alpha \epsilon_{abcde} \Sigma^{ab} \Sigma^{cd} \Phi^e
         = \alpha \left( q_3 H {\bar t}
                       + q_3 q_3 D 
                       + {\bar t} D {\bar \tau} \right),
\end{equation}
 where $\alpha$ is the coupling constant
 expected to be of the order of unity,
 $q_3$, ${\bar t}$ and $\bar \tau$ are chiral superfields
 which are contained in the field $\Sigma^{ab}$,
 and $H$ and $D$ are the chiral superfields
 which are contained in the field $\Phi^a$.
The first term of the above equation
 is the Yukawa coupling for the mass of the top quark.
The exact magnitude of the coupling $\alpha$ could be determined,
 if we have enough information of the K\"ahler potential
 for the effective fields.
If we consider the explicit $SU(5)$ breaking effect,
 the coupling constants for each three terms are not necessarily equal.

In the paper of Ref.\cite{K-O}
 the dynamical supersymmetry breaking,
 which is triggered by the strong $SU(2)_S$ supercolor dynamics,
 was introduced in this simple model described above.
It was shown that
 all gauginos (gluino, photino, wino, zino)
 and sfermions (squarks and sleptons) in third generation
 obtain their masses,
 which are consistent with the experiment,
 and the electroweak symmetry is broken
 due to the strong Yukawa coupling of the top quark
 through the radiative breaking mechanism\cite{radiative}.
But the origin of the masses of the bottom quark and tau lepton
 was not specified.

In this paper we proceed further
 by including other two generations
 and specifying the mechanism of the mass generation
 for all quarks and leptons.
In the next section
 we describe the particle contents and tree level superpotential,
 and explain how the strong gauge dynamics works.
In section \ref{sec:mass}
 the mechanism of the mass generation for quarks and leptons is described.
The Yukawa couplings for the masses of up-type quarks are generated
 by virtue of the non-perturbative gauge dynamics as explained above,
 and their hierarchy comes from the mixing
 between composite Higgs particles.
The Yukawa couplings
 for the masses of the bottom quark and tau lepton
 are generated through the exchange of a heavy particle
 which has strong relation with the dynamics of supersymmetry breaking.
The mass of the heavy particle is expected to be generated dynamically.
The masses of the strange and down quarks are generated
 through the kinetic mixing between the up-type quarks
 which is generated by the exchange of heavy particles
 whose masses are also expected to be generated dynamically.
The flavor mixing in the up-type quark sector
 is strongly related with the diagonal masses
 in the down-type quark sector in this model.
Unfortunately,
 the resultant masses of the strange and down quarks
 are too small to be realistic.
The masses of muon and electron are also generated in the same way,
 but they are also too small to be realistic.
In section \ref{sec:conclusion}
 we summarize the model, and describe some problems of this model.

\section{Three generation model}
\label{sec:three}

We introduce three generations
 as three copies of the same structure
 which is explained in the previous section.
Therefore,
 there are three $SU(2)$ hypercolor gauge symmetry for each generation,
 namely, $SU(2)_1$, $SU(2)_2$ and $SU(2)_H$
 for first, second and third generation, respectively.
In addition to that,
 the supercolor gauge interaction, $SU(2)_S$, is introduced,
 by which supersymmetry is dynamically broken
 through a mechanism which was proposed by Izawa et al.\cite{I-Y}.
The scales of dynamics of each gauge interactions are assumed as
\begin{equation}
 \Lambda_{(1)} \gg \Lambda_{(2)} \gg \Lambda_S > \Lambda_H.
\label{scales}
\end{equation}
We take $\Lambda_S \simeq 10^{10}$ GeV and $\Lambda_H \simeq 10^9$ GeV
 following the analysis of Ref.\cite{K-O}.

The particle contents of the model is as follows.
\begin{center}
\begin{tabular}{ccccccc}
 & \ $SU(2)_1 \ $ & $ \ SU(2)_2 \ $ & $ \ SU(2)_H \ $ &
   \ $SU(2)_S \ $ & $ \ SU(5) \ $ & \ $Z_2$ \ \\
$P^{(1)}$ & $2$ & $1$ & $1$ & $1$ & $5$ & $+$ \\
$N^{(1)}, N^{(1)}_i$ & $2$ & $1$ & $1$ & $1$ & $1$ & $-$ \\
$P^{(2)}$ & $1$ & $2$ & $1$ & $1$ & $5$ & $+$ \\
$N^{(2)}, N^{(2)}_i$ & $1$ & $2$ & $1$ & $1$ & $1$ & $-$ \\
$P$ & $1$ & $1$ & $2$ & $1$ & $5$ & $+$ \\
$N, N_i$ & $1$ & $1$ & $2$ & $1$ & $1$ & $-$ \\
${\bar \Phi}_A$ / ${\bar \Phi}_{3+A}$ &
 $1$ & $1$ & $1$ & $1$ & $5^*$ & $-$ / $+$ \\
$Q$ & $1$ & $1$ & $2$ & $2$ & $1$ & $+$ \\
${\tilde Q}_1$ / ${\tilde Q}_2$ &
 $1$ & $1$ & $1$ & $2$ & $1$ & $-$ / $+$ \\
$Z_1$ / $Z_2$ & $1$ & $1$ & $2$ & $1$ & $1$ & $-$ / $+$ \\
$Z$, $X$ / $Z'$, $X^{(A)}$ & $1$ & $1$ & $1$ & $1$ & $1$ & $-$ / $+$ \\
\end{tabular}
\end{center}
Here, $i=1,2$ and $A=1,2,3$.
The discrete symmetry distinguishes
 six \underline{5}${}^*$ multiplets in $SU(5)$
 into three Higgs multiplets and three matter multiplets
 which include right-handed down-type quarks
 and left-handed lepton weak doublets.
We introduce the following tree level superpotential
 which is consistent with the symmetry.
\begin{eqnarray}
 W_{tree} =&&
  \left(
   \begin{array}{ccc}
    [P^a N] & [P^{(2)a} N^{(2)}] & [P^{(1)a} N^{(1)}]
   \end{array}
  \right)
  \eta
  \left(
   \begin{array}{c}
    {\bar \Phi}_{3a} \\
    {\bar \Phi}_{2a} \\
    {\bar \Phi}_{1a}
   \end{array}
  \right)
 \nonumber\\
 && + \ [P^a N_i] \ \ \kappa^{(3)}_{iA} \ {\bar \Phi}_{Aa}
    + \ [P^{(2)a} N^{(2)}_i] \ \ \kappa^{(2)}_{iA} \ {\bar \Phi}_{Aa}
    + \ [P^{(1)a} N^{(1)}_i] \ \ \kappa^{(1)}_{iA} \ {\bar \Phi}_{Aa}
 \nonumber\\
 && + \lambda^{(3)}_{Z'} \ Z' \ [N_1 N_2]
    + \lambda^{(2)}_{Z'} \ Z' \ [N^{(2)}_1 N^{(2)}_2]
    + \lambda^{(1)}_{Z'} \ Z' \ [N^{(1)}_1 N^{(1)}_2]
 \nonumber\\
 && + \lambda^{(3)}_{X} \ X^{(3)} \ [N_1 N_2]
    + \lambda^{(2)}_{X} \ X^{(2)} \ [N^{(2)}_1 N^{(2)}_2]
    + \lambda^{(1)}_{X} \ X^{(1)} \ [N^{(1)}_1 N^{(1)}_2]
 \nonumber\\
 && + \kappa_{1A} \ [P^a Z_1] \ {\bar \Phi}_{Aa}
    + \kappa_{2A} \ [P^a Z_2] \ {\bar \Phi}_{3+A \, a}
 \nonumber\\
 && + \lambda_Z \ Z \ [Z_1 Z_2]
    + \lambda_X \ X \ [Z_1 Z_2]
  \nonumber\\
 && + \lambda \ Z \ [{\tilde Q}_1 {\tilde Q}_2]_S
    + \lambda' \ Z' \ [Q_1 Q_2]_S
    + \lambda_i \ [Z_i \ [Q {\tilde Q}_i]_S],
\label{tree}
\end{eqnarray}
 where square brackets
 mean the contraction of the indexes of hypercolor $SU(2)$ gauge groups
 (see Eq.(\ref{contraction})),
 and square brackets with subscript $S$
 mean the contraction of the indexes
 of the supercolor $SU(2)_S$ gauge group.
For simplicity,
 we do not include the explicit $SU(5)$ breaking effect 
 which can be easily incorporated whenever we want.
Unfortunately,
 this superpotential is not the general form in the given symmetry.
Although the unwanted interactions, like $ZZZ$ and $XXX$, are forbidden,
 several interactions, like $Z'Z'Z'$ and $Z[N Z_2]$,
 are dropped by hand.
This may suggest the additional symmetry
 or the modification of the dynamics of the supersymmetry breaking.
Since $U(1)_R$ symmetry is explicitly broken by gauge anomaly,
 there is no R-axion problem.
In the following,
 we regard all the coupling constants as the real ones, for simplicity.

Consider the confinement of the $SU(2)_S$ gauge interaction
 and the supersymmetry breaking.
Since there are four doublets of $SU(2)_S$,
 the low energy effective fields are expected as
\begin{equation}
 \left(
  \begin{array}{cc}
   (V+V') \epsilon_{\alpha\beta} & V_{j\alpha} \\
   -V_{i\beta}                   & (V-V') \epsilon_{ij}
  \end{array}
 \right)
 \sim
 \left(
  \begin{array}{cc}
   {[Q_\alpha Q_\beta]_S} & {[Q_\alpha {\tilde Q}_j]_S} \\
   {[{\tilde Q}_i Q_\beta]_S} & {[{\tilde Q}_i {\tilde Q}_j]_S}
  \end{array}
 \right),
\end{equation}
 with the constraint
\begin{equation}
 V^2 - V'^2 - [V_1 V_2] = \Lambda_S^4,
\end{equation}
 where the effective fields have mass dimension two\cite{S}.
Under the condition that
 the Yukawa couplings $\lambda$, $\lambda'$ and $\lambda_i$
 can be treated perturbatively,
 it is natural that the effective field
 $V \sim {1 \over 4} \{ \epsilon^{\alpha\beta} [Q_\alpha Q_\beta]
                      + \epsilon^{ij} [{\tilde Q}_i {\tilde Q}_j] \}$
 has vacuum expectation value due to the constraint.
Namely,
\begin{equation}
 V = \pm \Lambda_S^2
     \sqrt{1 + {{V'^2} \over {\Lambda_S^4}} 
             + {{[V_1 V_2]} \over {\Lambda_S^4}}}
     \longrightarrow
     \Lambda_S^2 + {1 \over 2} V'^2 + {1 \over 2} [V_1 V_2].
\end{equation}
Here, we rescaled the fields $V'$ are $V_i$
 by $\Lambda_S$ to have dimension one,
 and expanded by assuming
 that $V'/\Lambda_S$ and $V_i/\Lambda_S$ are small.
(We take positive sign by using the anomalous $U(1)_R$ symmetry.)
Then the superpotential of Eq.(\ref{tree}) effectively becomes
\begin{eqnarray}
 W_{eff} \simeq&&
  \left(
   \begin{array}{ccc}
    [P^a N] & [P^{(2)a} N^{(2)}] & [P^{(1)a} N^{(1)}]
   \end{array}
  \right)
  \eta
  \left(
   \begin{array}{c}
    {\bar \Phi}_{3a} \\
    {\bar \Phi}_{2a} \\
    {\bar \Phi}_{1a}
   \end{array}
  \right)
 \nonumber\\
 && + \ [P^a N_i] \ \ \kappa^{(3)}_{iA} \ {\bar \Phi}_{Aa}
    + \ [P^{(2)a} N^{(2)}_i] \ \ \kappa^{(2)}_{iA} \ {\bar \Phi}_{Aa}
    + \ [P^{(1)a} N^{(1)}_i] \ \ \kappa^{(1)}_{iA} \ {\bar \Phi}_{Aa}
 \nonumber\\
 && + \lambda^{(3)}_{Z'} \ Z' \ [N_1 N_2]
    + \lambda^{(2)}_{Z'} \ Z' \ [N^{(2)}_1 N^{(2)}_2]
    + \lambda^{(1)}_{Z'} \ Z' \ [N^{(1)}_1 N^{(1)}_2]
 \nonumber\\
 && + \lambda^{(3)}_{X} \ X^{(3)} \ [N_1 N_2]
    + \lambda^{(2)}_{X} \ X^{(2)} \ [N^{(2)}_1 N^{(2)}_2]
    + \lambda^{(1)}_{X} \ X^{(1)} \ [N^{(1)}_1 N^{(1)}_2]
 \nonumber\\
 && + \kappa_{1A} \ [P^a Z_1] \ {\bar \Phi}_{Aa}
    + \kappa_{2A} \ [P^a Z_2] \ {\bar \Phi}_{3+A \, a}
 \nonumber\\
 && + \lambda_Z \ Z \ [Z_1 Z_2]
    + \lambda_X \ X \ [Z_1 Z_2]
  \nonumber\\
 && + \lambda \ Z
      \left\{ \Lambda_S^2
            + {1 \over 2} V'^2 + {1 \over 2} [V_1 V_2]
            - \Lambda_S V' \right\}
  \nonumber\\
 && + \lambda' \ Z'
      \left\{ \Lambda_S^2
            + {1 \over 2} V'^2 + {1 \over 2} [V_1 V_2]
            + \Lambda_S V' \right\}
 \nonumber\\
 && + \lambda_i \Lambda_S [Z_i V_i].
\end{eqnarray}
Here, we dare to leave the fields
 which couple with $SU(2)_1$ and $SU(2)_2$,
 although they are confined at the scale $\Lambda_S$.
It can be shown that
 the $F$-component of $Z$ and $Z'$ have vacuum expectation values as
\begin{equation}
 \langle F_{Z} \rangle = - \lambda \Lambda_S^2,
 \quad
 \langle F_{Z'} \rangle = - \lambda \Lambda_S^2,
\end{equation}
 and supersymmetry is spontaneously broken,
 where we set $\lambda'=\lambda$, for simplicity.
We assume that vacuum expectation values
 of the scalar components of $Z_i$ and $V_i$,
 which parameterize the pseudo-flat direction,
 are fixed to zero by the effect of the K\"ahler potential.
(Note that
 since both $Z_i$ and $V_i$
 have the charge of $SU(2)_H$ gauge interaction,
 they should follow some non-trivial scalar potential
 which comes from the K\"ahler potential.)
The vacuum expectation values of the scalar components of
 $Z$, $Z'$, $X$ and $X^{(A)}$
 are also the parameters of the pseudo-flat direction.
We simply expect that
 these fields have vacuum expectation values 
 of the order of $10^{14}$ GeV with $\lambda, \ \lambda' \simeq 10^{-2}$
 through the effect of the K\"ahler potential
 (see Refs.\cite{K-O} and \cite{W}).
Once these assumptions are satisfied,
 the mechanism of the mediation of the supersymmetry breaking
 which is proposed by Ref.\cite{K-O} works.
All gauginos in the standard model have their masses
 which are compatible with experiments,
 and composite squarks and sleptons in the third generation
 have huge masses.
The elementary squarks and sleptons in third generation
 and the squarks and sleptons in first and second generations
 have their masses through the radiative correction
 in the same way in the gauge-mediated
 supersymmetry breaking model\cite{D-N}.

If the scalar component of $Z'$ and/or $X^{(A)}$
 have vacuum expectation values,
 fields $N_i$, $N^{(2)}_i$ and $N^{(1)}_i$ obtain their masses,
 and these fields can not be the component
 of massless composite fields at low energy.
We expect that
 even if the masses of $N_i^{(1)}$ and $N_i^{(2)}$ are smaller
 than the scales $\Lambda_{(1)}$ and $\Lambda_{(2)}$, respectively,
 they decouple from the low energy physics.
We integrate out these fields by using the conditions
 $\partial W_{eff} / \partial N_i = 0$
 and the same conditions for $N_i^{(1)}$ and $N_i^{(2)}$.
The same procedure can be applied for the field $Z_i$.
The field $Z_i$ has mass,
 if $Z$ and/or $X$ have vacuum expectation values.
The effective superpotential
 at the scale between $\Lambda_H$ and $\Lambda_S$ is obtained as
\begin{eqnarray}
 W_{eff} \simeq&&
  \left(
   \begin{array}{ccc}
    [P^a N] & \Lambda_{(2)} \Phi_2^a & \Lambda_{(1)} \Phi_1^a
   \end{array}
  \right)
  \left(
  \begin{array}{ccc}
    \eta_{33} & 0 & 0 \\
    \eta_{23} & \eta_{22} & 0 \\
    \eta_{13} & \eta_{12} & \eta_{11}
  \end{array}
  \right)
  \left(
   \begin{array}{c}
    {\bar \Phi}_{3a} \\
    {\bar \Phi}_{2a} \\
    {\bar \Phi}_{1a}
   \end{array}
  \right)
 \nonumber\\
 && - {{\kappa^{(1)}_{1A} \kappa^{(1)}_{2B} \Lambda_{(1)}}
       \over {\lambda^{(1)}_{Z'} \langle Z' \rangle
              + \lambda^{(1)}_{X} \langle X^{(1)} \rangle}}
      {\bar \Phi}_{Aa} \Sigma_1^{ab} {\bar \Phi}_{Bb}
    - {{\kappa^{(2)}_{1A} \kappa^{(2)}_{2B} \Lambda_{(2)}}
       \over {\lambda^{(2)}_{Z'} \langle Z' \rangle
              + \lambda^{(2)}_{X} \langle X^{(2)} \rangle}}
      {\bar \Phi}_{Aa} \Sigma_2^{ab} {\bar \Phi}_{Bb}
 \nonumber\\
 && - {{\kappa^{(3)}_{1A} \kappa^{(3)}_{2B}}
       \over {\lambda^{(3)}_{Z'} \langle Z' \rangle
              + \lambda^{(3)}_{X} \langle X^{(3)} \rangle}}
      {\bar \Phi}_{Aa} [P^a P^b] {\bar \Phi}_{Bb}
    - {{\kappa_{1A} \kappa_{2B}}
       \over {\lambda_Z \langle Z \rangle + \lambda_X \langle X \rangle}}
      {\bar \Phi}_{Aa} [P^a P^b] {\bar \Phi}_{3+B \, b}
 \nonumber\\
 && + {{\alpha^{(1)}} \over {2^3}}
      \epsilon_{abcde} \Sigma_1^{ab} \Sigma_1^{cd} \Phi_1^e
    + {{\alpha^{(2)}} \over {2^3}}
      \epsilon_{abcde} \Sigma_2^{ab} \Sigma_2^{cd} \Phi_2^e,
\end{eqnarray}
 where we neglect the supersymmetry breaking terms,
 for simplicity (see Ref.\cite{K-O} for the supersymmetry breaking).
Here, we introduced the low energy effective fields
 $\Phi_1 \sim [P^{(1)}N^{(1)}] $, $\Sigma_1 \sim [P^{(1)} P^{(1)}]$,
 $\Phi_2 \sim [P^{(2)}N^{(2)}]$ and $\Sigma_2 \sim [P^{(2)} P^{(2)}]$
 which interact with each other
 through the dynamically-generated Yukawa interaction
 of the last line of the above superpotential.
The form of the coupling matrix $\eta$ in the above superpotential
 is the general one under the global $SU(3)$ rotation
 on ${\bar \Phi}_{Aa}$ fields.
 
The confinement
 of $SU(2)_H$ occurs in succession,
 and the effective superpotential below the scale $\Lambda_H$
 is obtained as follows.
\begin{eqnarray}
 W_{eff} \simeq&&
      \sum_{a=1,2,3} \Phi_{\tilde A}^a M^D_{{\tilde A}{\tilde B}}
                   {\bar \Phi}_{{\tilde B}a}
    + \sum_{a=4,5} \Phi_{\tilde A}^a M^H_{{\tilde A}{\tilde B}}
                     {\bar \Phi}_{{\tilde B}a}
 \nonumber\\
 && - {{\kappa^{(1)}_{1A} \kappa^{(1)}_{2B} \Lambda_{(1)}}
       \over {\lambda^{(1)}_{Z'} \langle Z' \rangle
              + \lambda^{(1)}_{X} \langle X^{(1)} \rangle}}
      {\bar \Phi}_{Aa} \Sigma_1^{ab} {\bar \Phi}_{Bb}
    - {{\kappa^{(2)}_{1A} \kappa^{(2)}_{2B} \Lambda_{(2)}}
       \over {\lambda^{(2)}_{Z'} \langle Z' \rangle
              + \lambda^{(2)}_{X} \langle X^{(2)} \rangle}}
      {\bar \Phi}_{Aa} \Sigma_2^{ab} {\bar \Phi}_{Bb}
 \nonumber\\
 && - {{\kappa^{(3)}_{1A} \kappa^{(3)}_{2B} \Lambda_H}
       \over {\lambda^{(3)}_{Z'} \langle Z' \rangle
              + \lambda^{(3)}_{X} \langle X^{(3)} \rangle}}
      {\bar \Phi}_{Aa} \Sigma_3^{ab} {\bar \Phi}_{Bb}
    - {{\kappa_{1A} \kappa_{2B} \Lambda_H}
       \over {\lambda_Z \langle Z \rangle + \lambda_X \langle X \rangle}}
      {\bar \Phi}_{Aa} \Sigma_3^{ab} {\bar \Phi}_{3+B \, b}
 \nonumber\\
 && + {{\alpha^{(1)}} \over {2^3}}
      \epsilon_{abcde} \Sigma_1^{ab} \Sigma_1^{cd} \Phi_1^e
    + {{\alpha^{(2)}} \over {2^3}}
      \epsilon_{abcde} \Sigma_2^{ab} \Sigma_2^{cd} \Phi_2^e
    + {{\alpha^{(3)}} \over {2^3}}
      \epsilon_{abcde} \Sigma_3^{ab} \Sigma_3^{cd} \Phi_3^e,
\label{effective-W}
\end{eqnarray}
 where we introduced the low energy effective fields
 $\Phi_3 \sim [P N] $ and $\Sigma_3 \sim [P P]$
 which interact with each other through the dynamically-generated
 Yukawa interaction of the last term of the last line.
The explicit $SU(5)$ breaking is considered
 in the first line of the above superpotential
 (the coupling matrix $\eta$
 is decomposed to $\eta^D$ for $a=1,2,3$ and $\eta^H$ for $a=4,5$),
 and the first and second terms in the first line are the mass terms
 for colored Higgs and Higgs fields, respectively.
(Note that the indexes ${\tilde A}$ and ${\tilde B}$
  run reverse way, namely, from $3$ to $1$.)
The form of these mass matrixes are
\begin{equation}
 M^D \simeq
 \left(
 \begin{array}{ccc}
  \eta^D_{33} \Lambda_H & 0 & 0 \\
  \eta^D_{23} \Lambda_{(2)} & \eta^D_{22} \Lambda_{(2)} & 0 \\
  \eta^D_{13} \Lambda_{(1)} & \eta^D_{12} \Lambda_{(1)} &
  \eta^D_{11} \Lambda_{(1)}
 \end{array}
 \right),
 \qquad
 M^H \simeq
 \left(
 \begin{array}{ccc}
  \eta^H_{33} \Lambda_H & 0 & 0 \\
  \eta^H_{23} \Lambda_{(2)} & \eta^H_{22} \Lambda_{(2)} & 0 \\
  \eta^H_{13} \Lambda_{(1)} & \eta^H_{12} \Lambda_{(1)} &
  \eta^H_{11} \Lambda_{(1)}
 \end{array}
 \right).
\label{higgs-mass}
\end{equation}
We assume that
 the matrix elements of $\eta^H$
 are the same order of magnitude, except for 33-element.
We also assume that
 the matrix elements of $\eta^D$
 are the same order of magnitude, except for 33-element.
The values of $\eta^H_{33} \Lambda_H \equiv \mu \simeq 100$ GeV,
 namely $\eta^H_{33} \simeq 10^{-7}$,
 and $\eta^D_{33} \Lambda_H \equiv \mu_D \simeq 1000$ GeV,
 namely $\eta^D_{33} \simeq 10^{-6}$
 are necessary to have the electroweak symmetry breaking
 through the radiative breaking mechanism\cite{K-O}.
The eigenvalues of these mass matrixes are of the order of
 $\mu$, $\rho_{(2)} \sim \eta^H_{{\tilde A}{\tilde B}} \Lambda_{(2)}$,
 $\rho_{(1)} \sim \eta^H_{{\tilde A}{\tilde B}} \Lambda_{(1)}$
 for $M^H$ and 
 $\mu_D$, $\rho^D_{(2)} \sim \eta^D_{{\tilde A}{\tilde B}} \Lambda_{(2)}$,
 $\rho^D_{(1)} \sim \eta^D_{{\tilde A}{\tilde B}} \Lambda_{(1)}$
 for $M^D$ (${\tilde A}, \ {\tilde B} \ne 3$).
The mass of the colored Higgs
 which couples with the first generation particles
 through the dynamically-generated Yukawa interactions
 must be larger than $10^{17}$ GeV not to have rapid proton decay.
Therefore, we assume $\rho^D_{(1)} \simeq 10^{17}$ GeV,
 namely $\eta^D_{{\tilde A}{\tilde B}} \simeq 1$
 (${\tilde A}, \ {\tilde B} \ne 3$)
 and $\Lambda_{(1)} \simeq 10^{17}$.
This huge hierarchy in the coupling matrix $\eta$
 is one of the problems of this model.

\section{Mass generation mechanism}
\label{sec:mass}

First, we describe the generation of the mass hierarchy of up-type quarks.
The interactions of the last line of Eq.(\ref{effective-W})
 contain the Yukawa couplings for the masses of up, charm and top quarks,
 but there are three pairs of Higgs doublets in each generation.
The Higgs fields in each generation mix with each other
 through the mass matrix $M^H$.
The mass matrix is approximately diagonalized as
 $U M^H V^{\dag} \simeq \mbox{\rm diag}(\mu \ \rho_{(2)} \ \rho_{(1)})$,
 where the order of magnitude of the matrix elements of $U$ is
\begin{equation}
 U \sim
    \left(
     \begin{array}{ccc}
      1 & {\mu \over {\rho_{(2)}}} & {\mu \over {\rho_{(1)}}} \\
      {\mu \over {\rho_{(2)}}} & 1 & {{\rho_{(2)}} \over {\rho_{(1)}}} \\
      {\mu \over {\rho_{(1)}}} & {{\rho_{(2)}} \over {\rho_{(1)}}} & 1
     \end{array}
    \right),
\label{diagonalize}
\end{equation}
 and all the elements of the matrix $V$ are of the order of unity.
According to Eq.(\ref{scales}),
 the hierarchy
 $\mu/\rho_{(2)}, \mu/\rho_{(1)}, \rho_{(2)}/\rho_{(1)} \ll 1$
 is assumed.
We can identify the lightest Higgs pair
 to the Higgs pair in the minimal supersymmetric standard model.
The other Higgs pairs are heavy and decouple from the low energy physics.
Therefore, Yukawa couplings for up-type quarks can be described as
\begin{equation}
 W_Y^{up} \simeq
      {{\alpha^{(1)}} \over {2^3}} {\mu \over {\rho_{(1)}}}
      \epsilon_{abcde} \Sigma_1^{ab} \Sigma_1^{cd} \Phi^e
    + {{\alpha^{(2)}} \over {2^3}} {\mu \over {\rho_{(2)}}}
      \epsilon_{abcde} \Sigma_2^{ab} \Sigma_2^{cd} \Phi^e
    + {{\alpha^{(3)}} \over {2^3}}
      \epsilon_{abcde} \Sigma_3^{ab} \Sigma_3^{cd} \Phi^e,
\label{yukawa-dynamical}
\end{equation}
 where $\Phi$ denotes the lightest Higgs multiplet,
 and the index $e$ takes the values only $4$ and $5$.
Then we have relations
\begin{equation}
 {{m_c} \over {m_t}} \simeq {\mu \over {\rho_{(2)}}},
 \qquad
 {{m_u} \over {m_t}} \simeq {\mu \over {\rho_{(1)}}}.
\end{equation}
Since we need to take $\mu \simeq 100$GeV
 for the electroweak symmetry breaking,
 we have $\rho_{(2)} \simeq 10^4$ GeV and $\rho_{(1)} \simeq 10^7$ GeV,
 namely $\eta^H_{{\tilde A}{\tilde B}} \simeq 10^{-10}$
 (${\tilde A}, \ {\tilde B} \ne 3$)
 and $\Lambda_{(2)} \simeq 10^{14}$ GeV.
 
Next, we discuss the mass generation of down-type quarks.
The interaction of the last term
 in the third line of Eq.(\ref{effective-W}) 
 gives the following Yukawa couplings.
\begin{equation}
 {\cal L}_Y^{down} = 
  - {{\Lambda_H}
     \over {\lambda_Z \langle Z \rangle + \lambda_X \langle X \rangle}}
  \kappa
  \left\{ \kappa_{21} {\bar H} {\bar b}_R q_{3L}
        + \kappa_{22} {\bar H} {\bar s}_R q_{3L}
        + \kappa_{23} {\bar H} {\bar d}_R q_{3L}
  \right\}
  + \mbox{\rm h.c.},
\label{yukawa-Z}
\end{equation}
 where ${\bar H}$ is the lightest $SU(2)_L$-doublet Higgs scaler field,
 $b_R$, $s_R$ and $d_R$ are right-handed quark fermion fields,
 and $q_{3L}$ is the left-handed $SU(2)_L$ doublet quark fields
 in third generation.
We defined
 $\kappa \equiv \kappa_{13} V_{33}
              + \kappa_{12} V_{32} + \kappa_{11} V_{31}$,
 where $V_{33}$, $V_{32}$ and $V_{31}$
 are matrix elements of the matrix $V$.
It is clear that
 the Yukawa coupling for the bottom quark mass is included
 in Eq.(\ref{yukawa-Z}).
Since it is natural to take
 $\lambda_Z, \  \lambda_X \simeq \lambda \simeq 10^{-2}$
 and $\lambda_Z \langle Z \rangle
    + \lambda_X \langle X \rangle \simeq 10^{12}$,
 we have the Yukawa coupling for the bottom quark mass as
\begin{equation}
 g_b = \kappa \kappa_{21} 10^{-3}.
\end{equation}
It is expected that
 the Yukawa coupling for the top quark mass is of the order of unity
 and $\tan \beta \simeq 4$ (see Ref.\cite{K-O}).
Therefore, $g_b = (m_b / m_t) \tan \beta \simeq 0.1$
 and $\kappa \kappa_{21}$ must be of the order of $10^2$.
The mixing masses
 between $b_L$ and $s_R$ and between $b_L$ and $d_R$,
 which come from the second and third terms
 of Eq.(\ref{yukawa-Z}), respectively,
 are expected to be the same order of the bottom quark mass.
We can have the mass of the tau lepton
 which is the same order of the bottom quark mass.

The Yukawa couplings for the masses of the strange and down quarks
 are not included in the effective superpotential
 of Eq.(\ref{effective-W}).
However, if we have mixing in the up-type quark sector,
 the masses of these quarks can be generated by the quantum correction
 through the diagram of Fig.\ref{mass}.
Although there is no mass mixing in the up-type quark sector,
 the kinetic mixing is generated through the diagram of Fig.\ref{mixing}.

Consider the generation of the strange quark mass.
The kinetic mixing between top and charm quarks are estimated as follows.
\begin{equation}
 {\cal L}^{tc}
  = \varepsilon_{tc} \ {\bar t} i \gamma^\mu D_\mu c + \mbox{\rm h.c.}
\end{equation}
\begin{equation}
 \varepsilon_{tc} \simeq
  \left( {{\kappa^{(3)}_{13} \kappa^{(3)}_{22} \Lambda_H}
         \over {\lambda^{(3)}_{Z'} \langle Z' \rangle
              + \lambda^{(3)}_{X} \langle X^{(3)} \rangle}} \right)
  \left( {{\kappa^{(2)}_{13} \kappa^{(2)}_{22} \Lambda_{(2)}}
         \over {\lambda^{(2)}_{Z'} \langle Z' \rangle
              + \lambda^{(2)}_{X} \langle X^{(2)} \rangle}} \right)
  {1 \over {16 \pi^2}}
   \ln \left( {{\Lambda_H} \over {\rho_{(2)}}} \right)^2,
\end{equation}
 where we take $\Lambda_H$
 as the physical ultraviolet energy-momentum cut off in Euclidean space.
If we naturally take the values of Yukawa couplings as
 $\lambda^{(2),(3)}_{Z'},
  \lambda^{(2),(3)}_X \simeq \lambda \simeq 10^{-2}$,
 namely $\lambda_{Z'}^{(2),(3)} \langle Z' \rangle
    + \lambda_X^{(2),(3)} \langle X^{(2),(3)} \rangle \simeq 10^{12}$,
 and if we take $\kappa^{(2)}_{13} \kappa^{(2)}_{22}
                 \kappa^{(3)}_{13} \kappa^{(3)}_{22} \simeq 10$,
 we have $\varepsilon_{tc} \simeq 0.1$.
We assume that the perturbative calculation is good for order estimation,
 even if the coupling constant of the vertexes in Fig.\ref{mixing}
 is very large.

The diagram of Fig.\ref{mass} gives the strange quark mass as
\begin{equation}
 m_s \simeq  g_{s_R t_L} g_{c_R s_L} \varepsilon_{tc} \ 
             {{\sin\beta \cos\beta} \over {16 \pi^2}} \ 
             m_t \ln \left( {\Lambda_H \over {m_t}} \right)^2,  
\end{equation}
 where
\begin{equation}
 g_{s_R t_L}
  = {{\kappa \kappa_{22} \Lambda_H}
     \over {\lambda_Z \langle Z \rangle + \lambda_X \langle X \rangle}},
\quad
 g_{c_R s_L} \simeq {\mu \over {\rho_{(2)}}}
\end{equation}
 come from Eqs.(\ref{yukawa-Z}) and (\ref{yukawa-dynamical}),
 respectively.
The main contribution in the diagram of Fig.\ref{mass}
 is the would-be Nambu-Goldstone boson component
 of the Higgs fields in $R_\xi$-Landau gauge.
If we naturally take $g_{s_R t_L} \simeq g_b \simeq 0.1$,
 then we have $m_s \simeq 0.1$ MeV. 
This value is too small for the strange quark mass
 which is usually considered as $100 < m_s < 300$ MeV.
The reason of this small value
 is the smallness of
 $g_{c_R s_L} \simeq \mu / \rho_{(2)} \simeq 10^{-3}$
 and loop suppression factors of the diagrams.

The mass of the down quark can also be generated in the same way.
But the value is negligibly small,
 because the kinetic mixing coefficient $\varepsilon_{ut}$
 have to be vary small to have long life time of the proton
 ($\varepsilon_{ut} < 10^{-9}$).

The masses of muon and electron
 can be generated through the similar diagram of Fig.\ref{mass}
 in which Higgs fields are replaced by colored Higgs fields.
But the resultant masses are negligibly small, since
 the mixing angles between light colored Higgs and heavy colored Higgs
 are very small.

\section{Conclusion}
\label{sec:conclusion}

In this paper we proposed a supersymmetric composite model
 in which the both supersymmetry and electroweak symmetry
 are dynamically broken,
 and the mass generation mechanism for all quarks and leptons
 is explicitly described.
The particle contents are very simple.
Only the particles
 which belong to the fundamental representation
 of each simple unitary gauge group are considered.
Since the representation is vector-like in each gauge group,
 including the gauge group of the standard model,
 the anomaly cancelation is trivial.

The mass hierarchy in the up-type quark sector
 can be clearly understood as the result of the mixing
 between the composite Higgs of each generations.
The generation of the mass hierarchy
 for down-type quarks and charged leptons is more complicated.
The bottom quark and tau lepton are special particles,
 since they can directly couple
 with the dynamics of the supersymmetry breaking.
Therefore,
 they have relatively large masses
 in comparison with the masses of the corresponding particles
 in other generations.
The generation of the Yukawa couplings
 for the masses of down-type quarks and charged leptons
 in the other two generations are forbidden at the tree level
 by the discrete $Z_2$ symmetry
 which distinguishes the six multiplets
 in \underline{5}${}^*$ representation of $SU(5)$
 into three Higgs multiplets and three matter multiplets.
(This discrete symmetry also forbids the Yukawa coupling at tree level
 which contributes to the dimension-five operator\cite{dim-5}
 for proton decay.)
But since the discrete symmetry
 is spontaneously broken by the vacuum expectation value of $Z$ or $X$,
 their masses can be generated through the quantum correction.
Unfortunately,
 the resultant masses are too small to be realistic,
 because of the many suppression factors
 and the constraint from the long life time of proton.
But it must be stressed that
 no mass scale is introduced by hand.

There are many other open questions.

We put several assumptions on the dynamics.
The most serious one is the unspecified dynamics
 to have vacuum expectation values of the scalar components
 of the gauge singlet fields $Z$, $X$, $Z'$ and $X^{(A)}$.
Since they are parameters of the pseudo-flat direction,
 it could be possible that they have vacuum expectation values
 through the quantum correction to the K\"ahler potential\cite{W}.
It could also be true
 that there is more appropriate dynamics for the supersymmetry breaking
 in which such vacuum expectation values are naturally generated.

Other problem is that
 we have to consider the hierarchical Yukawa coupling
 in the tree level superpotential,
 especially in the coupling matrix $\eta$.
It looks like very artificial and brings conceptual difficulty,
 because we are pursuing the origin of the mass hierarchy.

Although
 the model which is proposed in this paper is not the perfect one,
 we believe that it is a primitive one which is worth developing further.

\acknowledgments

I would like to thank T.~Appelquist, N.~Okada and F.~Sannino
 for a careful reading of the manuscript.
This work was supported in part
 by U.S. Department of Energy under Contract No.~DE-FG02-92ER-40704
 and in part by Grant-in-Aid for Scientific Research
 on Priority Areas (Physics of CP violation)
 from the Ministry of Education, Science, and Culture of Japan.

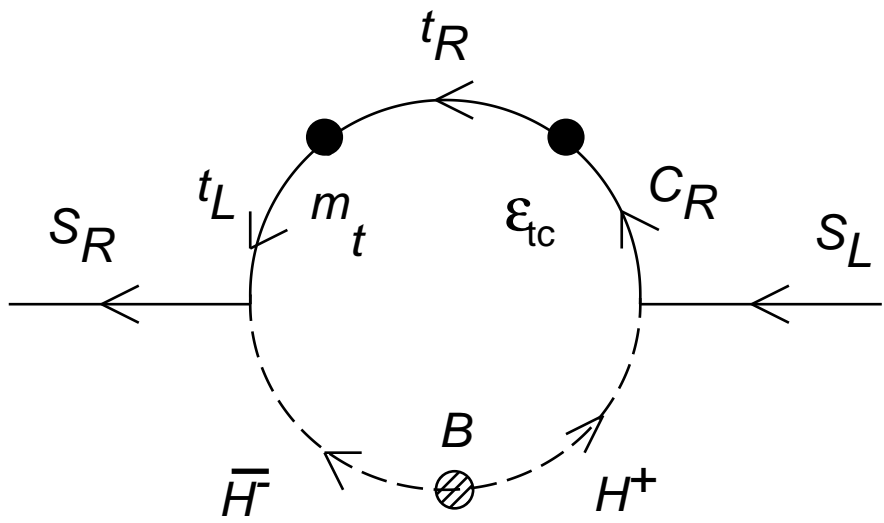
\begin{figure}
\caption{
The diagram for the strange quark mass.
The factor $\varepsilon_{tc}$ denotes the coefficient
 of the kinetic mixing between top and charm quarks,
 and $B$ denotes the supersymmetry breaking mass
 for the lightest Higgs field.
The diagram for the down quark mass
 is obtained by replacing the charm quark inside the loop
 with the up quark.}
\label{mass}
\end{figure}
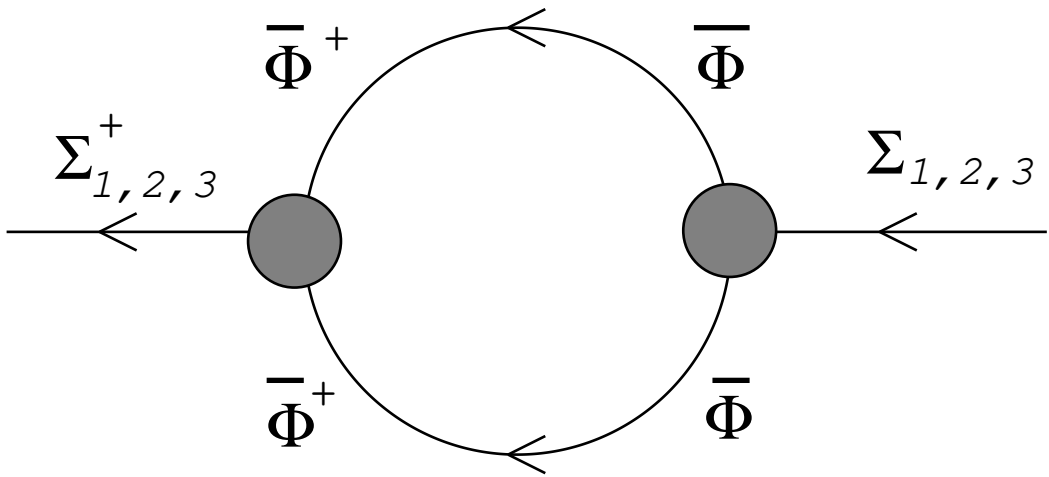
\begin{figure}
\caption{
The supergraph for the kinetic mixing between the up-type quarks.}
\label{mixing}
\end{figure}
\end{document}